# Normalization and selecting non-differentially expressed genes improve machine learning modelling of cross-platform transcriptomic data

Fei Deng[1], Catherine H Feng,[1,2], Nan Gao,[3,4] Lanjing Zhang[1, 4, 5, 6, *]

[1]Department of Chemical Biology, Ernest Mario School of Pharmacy, Rutgers University, Piscataway, NJ, USA.

[2] Harvard University, Cambridge, MA, USA

[3] Department of Biological Sciences, School of Arts & Sciences, Rutgers University, Newark, NJ, USA.

[4] Department of Pharmacology, Physiology, and Neuroscience, New Jersey Medical School, Rutgers University, Newark, NJ, USA.

[5] Department of Pathology, Princeton Medical Center, Plainsboro, NJ, USA.

[6] Rutgers Cancer Institute of New Jersey, New Brunswick, NJ, USA.

*Correspondence to: Lanjing Zhang, Department of Pathology, Princeton Medical Center, 1 Plainsboro Rd, Plainsboro, NJ 08536, USA. Fax: +609-853-6841, lanjing.zhang@rutgers.edu

## Abstract:

Normalization is a critical step in quantitative analyses of biological processes. Recent works show that cross-platform integration and normalization enable machine learning (ML) training on RNA microarray and RNA-seq data, but no independent datasets were used in their studies. Therefore, it is unclear how to improve ML modelling performance on independent RNA array and RNA-seq based datasets. Inspired by the house-keeping genes that are commonly used in experimental biology, this study tests the hypothesis that non-differentially expressed genes (NDEG) may improve normalization of transcriptomic data and subsequently cross-platform modelling performance of ML models. Microarray and RNA-seq datasets of the TCGA breast cancer were used as independent training and test datasets, respectively, to classify the molecular subtypes of breast cancer. NDEG ($p>0.85$) and differentially expressed genes (DEG, $p<0.05$) were selected based on the p values of ANOVA analysis and used for subsequent data normalization and classification, respectively. Models trained based on data from one platform were used for testing on the other platform. Our data show that NDEG and DEG gene selection could effectively improve the model classification performance. Normalization methods based on parametric statistical analysis were inferior to those based on nonparametric statistics. In this study, the LOG_QN and LOG_QNZ normalization methods combined with the neural network classification model seem to achieve better performance. Therefore, NDEG-based normalization appears useful for cross-platform testing on completely independent datasets. However, more studies are required to examine whether NDEG-based normalization can improve ML classification performance in other datasets and other omic data types.

## Introduction

Normalization is a critical step in quantitative analyses of biological processes, but very difficult yet important in cross-platform comparison. Independent dataset is required for rigorous testing of any quantitative biological analyses, while high-throughput transcriptomic data can be obtained using two different platforms, namely RNA microarray and recently RNA-sequencing (RNA-seq). The cross-platform difference makes direct cross-platform testing in an independent dataset challenging, if not impossible. Therefore, this study aimed to improve performance of machine learning (ML) modelling of transcriptomic data cross the two commonly used high-through RNA quantification platforms.

Advances in genome sequencing technology have given researchers a whole new perspective on fighting a variety of complex diseases.[1-3] Cancer is a complex genetic disease involving multiple subtypes. In order to better understand this disease to improve the accuracy and reliability of diagnosis, treatment and prognosis prediction, researchers have collected massive amounts of gene expression data in different biological environments and through different assays. Analyzing these data and mining the important relationship between them and the disease also puts brand new requirements on algorithms for data processing, prediction and classification.

How to rationally and adequately apply these data from different platforms, researchers have done a lot of work for this purpose. Most of them consider various ways to eliminate or reduce the data differences cross platforms, and then incorporate them into the same framework for analysis, which has the most direct benefit of expanding the amount of information used for analysis.[4-12] However, it also introduces selection biases by selecting/reducing features. Whereas we seek to unbiasedly normalize biological data, which it appears more complex yet more rigorous.

ML methods excel at solving complex problems such as tumor subtype classification, and often are trained using large amounts of data to find the hidden patterns needed to make decisions[4,13-17]. However, there are several key issues when classifying tumor subtypes based on gene expression data, such as high dimensionality and class imbalance[18-20]. High dimensionality of the data refers to the presence of an exceptionally large number of features, i.e., genes, compared to the samples. To address the high dimensionality problem, many feature selection methods and techniques have been devised to remove irrelevant features, reduce model training time, and develop generalized and scalable models.[19,21-28] These feature selection algorithms rely on optimization algorithms or statistical methods and are broadly classified into packing, hybrid and filtering methods. However, there is no generalized method that can handle omic datasets for all platforms. In addition, gene screening strategies play an important role in finding key genes such as housekeeping. Most studies have used software such as GeNorm, BestKeeper and NormFinder to analyze the expression stability of certain genes of interest in disease groups and healthy controls to identify reference genes. There have also been successes in identifying key genes through machine learning methods.[14,29-38]

Normalization method is another issue that has received a lot of attention. It can effectively address the problem of discrepancies between gene expression data from different samples obtained from different platforms: reducing discrepancies due to technical reasons, but retaining those due to biological reasons. The importance of data normalization for constructing predictive models based on machine learning algorithms has been validated in a large number of research efforts, which have also shown that differences in normalization methods have different impacts on different machine learning models.[4,8,15,39-56] However, when cross-platform genetic data analysis is performed, no study has yet delved into how to optimize tumor subtype classification models under the interplay between feature selection methods, normalization methods, and machine learning algorithms.

Here, we will propose a cross-platform data analysis method for tumor subtype prediction based on cross-platform genetic data through a series of experiments. We will study how to reasonably select stable genes for normalization and differential genes for classification when models trained on RNA-seq data are used for the prediction of microarray data or when models trained on microarray data are used for the prediction of RNA-seq data, and we will analyze which combined use of normalization methods and supervised machine learning methods can achieve better tumor subtype prediction in our Effect. Taking this study as an example, we hope to provide researchers with a comprehensive selection strategy for various classification prediction studies based on genetic data.

The structure of the paper is description of data used, elaboration of the proposed method, demonstration of results and comparison of models followed by Discussions.

## Dataset Description

To fulfill the experimental requirements, the datasets we chose had to have matched genes present on both microarrays and RNA-seq, and a sufficient number of labeled samples.

The Breast Cancer (BRCA) dataset from The Cancer Genome Atlas (TCGA) include samples measured with both microarray and RNA-seq platforms and well-defined molecular subtypes,

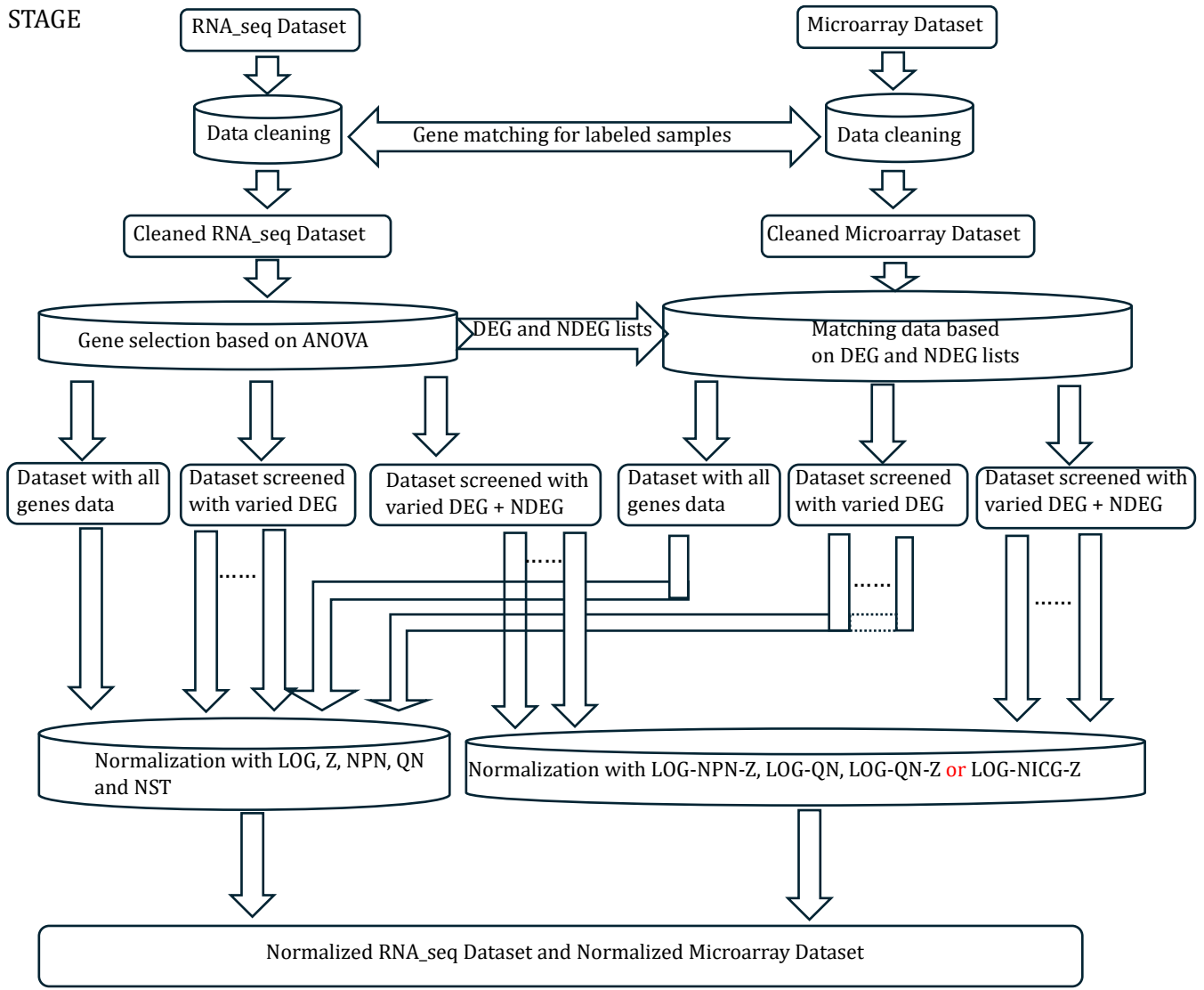

Figure 1: Stage 1 of the framework of the classification strategy: data cleaning, gene selection and normalization (RNAseq Dataset as training set and Microarray Dataset as testing set)

which are well suited to be used as class labels for supervised machine learning models. We restricted the datasets of both platforms to BRCA tumor samples with corresponding molecular subtype labels first. Thus, 520 samples were selected from 597 microarray samples, and 522 samples were selected from 1,215 RNA-seq samples, of which the microarray samples included 96 cases of Basal, 58 cases of Heritage, 231 cases of LumA, 127 cases of LumB, and 8 cases of Normal. The ratio of the number of samples in the largest class to the smallest class is approximated to be 29:1, which is a typical unbalanced dataset. These 520 microarray samples

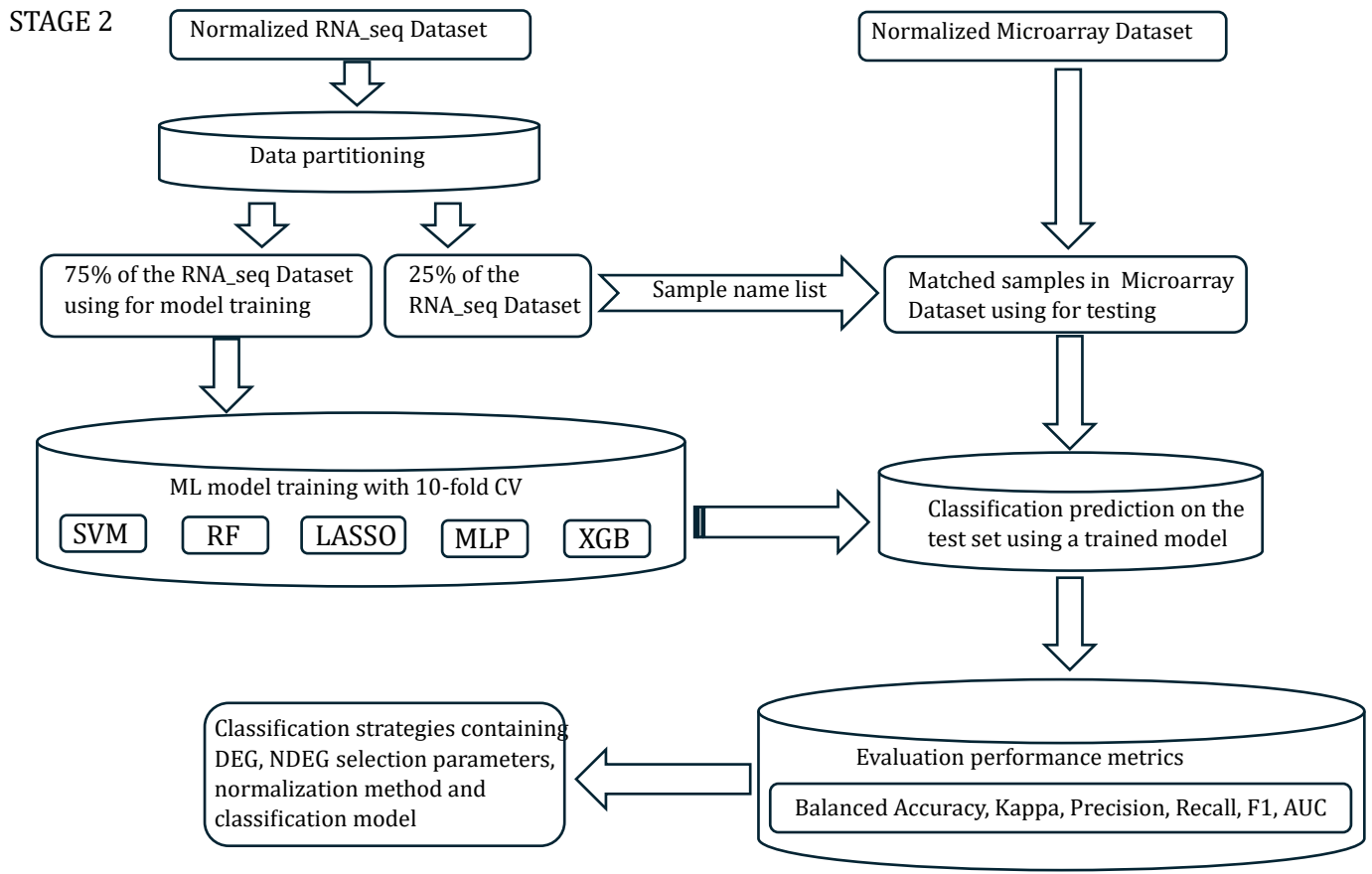

Figure 2: Stage 2 of the framework of the classification strategy: dataset partitioning, classification model training, prediction and classification performance evaluation (RNAseq Dataset as training set and Microarray Dataset as testing set)

exist in the 522 RNA-seq samples at the same time, and the RNA-seq platform has two more Basal samples.

## Method

A flowchart for training on RNA-seq data and testing on microarray data has been divided into two stages and shown in **Fig. 1** and **Fig. 2** (Supplementary **Fig. 1** and **Fig. 2** show a flowchart for training on microarray data and testing on RNA-seq data). The entire process was repeated at least five times to obtain a relatively comprehensive model assessment. The analysis steps of each process mainly include: data cleaning, gene selection, normalization, dataset partitioning, classification model training, prediction and classification performance evaluation. Python version 3.11.9 64-bit is used for the code implementation.

For the convenience of the subsequent narrative, we will refer to the model training based on the RNA-seq data and testing based on the microarray data as Model-S; and the model training based on the microarray data and testing based on the RNA-seq data as Model-A.

## Data cleaning

Samples were first screened against the data from both platforms, retaining only those samples with corresponding subtype classification labels. Genes that were present in the datasets of both platforms were retained by gene matching. Then the corresponding genes with missing values of gene expression values were removed. After data cleaning, only the expression values of 15,672 shared genes under the samples with classification labels were left in the datasets of both platforms.

## Gene Selection

In this study, the number of samples is significantly smaller than the number of features (15,672 genes). Many features may be redundant, which will lead to multicollinearity. Due to the high number of features relative to the number of samples, models are prone to overfitting. It is thus possible that the models fit the noises in the training data rather than the underlying pattern. In addition, there will be an increase in computational cost and decrease in interpretability.

Given the challenges posed by high-dimensional data, feature selection reduction is often required to improve model performance and interpretability. There are several common approaches for feature selection: filtering, wrapping and embedding methods. Here, we performed a one-way ANOVA, a filter method based on statistical analysis, on the data from each of the two platforms separately.

ANOVA is used to compare between-group variance (differences between category means) and within-group variance (fluctuations within the same category) for data sets with multiple categories to determine if at least one group's mean is significantly different from the others. The F-value is the ratio of two variances and represents the variance of the between-group means compared to the within-group variance. It is used to test the null hypothesis, which states that all group means are equal.[36,56,57]

The F-value in ANOVA is calculated as follows:

$$F = \frac{\text{MSB}}{\text{MSW}} = \frac{\frac{\sum_{i=1}^{k} n_i(\overline{Y_i}-\overline{Y})^2}{k-1}}{\frac{\sum_{i=1}^{k}\sum_{j=1}^{n_i}(Y_{ij}-\overline{Y_i})^2}{N-k}} \qquad (1)$$

where MSB (Mean Square Between-group) is the sum of squares between groups divided by *k-1*, the degrees of freedom between (number of categories minus one), and MSW (Mean Square Within) is the sum of squares within groups divided by *N-k*, the degrees of freedom within (total sample size minus the number of groups). $N$ is the total number of observations, $k$ is the number of groups, $n_i$ is the number of observations in group $i$, $\overline{Y_i}$ is the mean of group $i$, and $\overline{Y}$ is the overall mean.

A high F-value indicates a greater likelihood that the between-group variance is much larger than the within-group variance, suggesting that there are significant differences in means between groups. Genes that fulfill such condition are suitable for classification, which we call differential genes (DEG). Conversely, a low F-value indicates less significant differences. Such genes are suitable as reference genes for normalization, which we name non-differential genes (NDEG).

To follow statistical principles of gene selection, F-values are first calculated from gene expression data and sample category labels, then compared with the theoretical values in the F-distribution table to determine the P-value. The P-value represents the probability of observing the current or more extreme F-value under the null hypothesis (that all group means are equal). If the P-value is less than a preset significance level (e.g., 0.05), the null hypothesis is rejected, indicating that at least one group's mean is significantly different. By setting different thresholds, the corresponding gene sets can be determined. For example, when the threshold is 0.95, genes with P > 0.95 are selected as a set of NDEG for normalization. When the threshold is 0.05, genes with P < 0.05 are selected as a set of DEG for classification. The effects of different NDEG and DEG gene sets on the classification prediction results were observed by varying the thresholds.

## Data Partitioning

In order to fairly evaluate the prediction performance on data from one platform of a classification model trained on data from another platform, a rational dataset partitioning strategy needs to be designed. Repeated validation and hold-out methods are two commonly used methods for machine learning model evaluation. Repeated validation refers to evaluating the performance of a model multiple times using different training and test sets, and then taking the average as the final performance estimate. The hold-out method, on the other hand, pre-divides a portion of the dataset as a test set, then uses the training set to train the model and the test set to evaluate the model's performance. On a small dataset, holding a larger percentage of data for testing may result in insufficient training data, which may affect the model performance, while holding a smaller percentage of data may lead to unstable results, as some important features may not be adequately represented in the test set.

Therefore, we adopt a repeated validation approach based on bootstrap technique here to evaluate the model performance. When we randomly select some samples on the RNA-seq platform for training, then the remaining samples in the microarray dataset that do not overlap with these samples are used for testing, and vice versa. To ensure the fairness of the model evaluation, during the completion of the complete round of analysis shown in the flowchart, the samples constituting the training data and the test set were kept constant throughout the process, regardless of how the gene selection thresholds were varied and how the normalization methods and classifiers were combined. The training data are further randomly divided into training and validation sets to complete the training of the mode.

Since the data itself has five categories of cancer subtypes with very large non-equilibrium, the data will be divided into training data and test set in the ratio of 75:25 while maintaining the original category ratio. Under Model-S, 75% (390 samples) of the 522 RNA-seq samples were randomly selected. To find the best performing model configuration, the validation was done by $k$-fold cross validation technique with K value considered to be 10. After the training was completed, the samples with the same names as the samples involved in the training were removed from the Microarray data and only the remaining 131 samples that do not overlap constitute the test set for performance evaluation. In Mode-A, 75% (389 samples) of the 520 Microarray data samples were randomly selected to form the training and validation sets, while the corresponding samples in the RNA-seq data were removed, and only the remaining 133 samples that did not overlap were retained to form the test set. When randomly dividing the training data and test set, the proportion of the number of samples in each category was always kept the same as in the raw data set.

## Normalization

Among the main steps in the processing of genetic data, normalization is essential and its importance is well recognized. There are many normalization methods, and the choice of which method to use is related to the data and the goal of processing. Here we choose only a few commonly used normalization methods for comparative analysis to refine our processing strategy.

We first investigated the effect of five commonly used normalization methods on data preprocessing on both the full gene data and data screened with DEG genomes selected with different thresholds. These methods include LOG, Z, NPN, QN, and NST. We then investigated the effects of four reference gene-based normalization methods, including LOG-NPN-Z, LOG-QN, LOG-QN-Z, and LOG-NICG-Z.

After applying the same normalization methods to the training data and test set, different classification learning models are used for training and testing. The impact of different normalization methods during data analysis was evaluated by comparing these results with that of direct classification prediction on the original data.

### *Log2-transformation (LOG)[4,15]*

Genetic data usually has a large dynamic range and skewed distribution. Logarithmic transformation helps to reduce the dynamic range and skewness, enhance symmetry and normality, and better satisfy downstream statistical analysis assumptions. To solve the zero-value problem, a small constant (e.g., 1) is added before the logarithmic transformation.

### *Z-Score Transformation (Z)[4,15]*

Z-Score transformation is a commonly used method of normalizing data to a standard normal distribution and making different features comparable. First, the mean ($\mu$) and standard deviation ($\sigma$) are calculated for each genetic trait, and then the data points for each trait are normalized using $(x-\mu)/\sigma$, resulting in a mean of 0 and a standard deviation of 1 for each trait.

*Normal Score Transformation (NST)[58]*

Normal Score Transformation is a technique used to convert ranks of data into their standard normally distributed counterparts. This transformation ensures that the data follow a normal distribution and therefore helps to generate hypotheses for certain statistical analyses that assume normality, such as linear regression and analysis of variance. It first ranks the data in ascending order and then converts the ranks into percentiles, which represent the position of each data point in the cumulative distribution function (CDF) of the data. Finally, these percentiles are mapped to z-scores using the inverse CDF (quantile function) of the standard normal distribution.

*Non-parametric Normalization (NPN)[59]*

Non-parametric normalization (NPN) is a technique commonly used to normalize high-dimensional data, such as gene expression data, by transforming it to make genetic data more comparable across samples or experimental conditions and to reduce technical variability while preserving the biological signal. Unlike parametric methods, which assume a particular data distribution (e.g., normal distribution), non-parametric normalization does not assume a specific distribution and are therefore more robust to a wide range of data types and distributions. Non-parametric normalization works by mapping observations to a typical distribution (e.g., normal) using the rank or order of magnitude of the data. First, the data are categorized and ranks are assigned to each sample. The rank is then converted to a percentile, which is the percentile of each data point calculated from the rank to represent the relative position of a value in the data set. Finally, the percentile is then mapped to a typical distribution or other reference distribution.

While both NST and NPN involve ranking data and mapping it to a reference distribution, NST more specifically transforms data to a normal distribution for statistical analysis, while NPN is a more flexible approach for normalizing high-dimensional data without assuming a specific data distribution.

*Quantile Normalization (QN) [4]*

QN that do not depend on a specific reference gene assume that most genes are expressed at similar levels across samples. QN can match the expression distribution across all samples by aligning the quantitative values for each sample, thereby reducing outliers or extreme expression values, minimizing technical variability, and improving sample comparability. If biological differences are the primary cause of data variability, QN can capture and preserve these differences more effectively by adjusting the overall distribution. Specifically, QN is implemented by sorting the data and replacing each sample value for a given class with the mean of all samples in that class.

QN by partial reference gene is a specific normalization method commonly used for gene expression data, especially in high-throughput analyses such as microarrays or RNA-seq. A set of genes that are assumed to be stably expressed under different conditions (i.e., housekeeping genes) is used to adjust the expression levels of other genes, thereby reducing technical variability and improving data comparability. When the screened NDEG genes are used for QN, only the expression of each sample on the NDEG genes is selected to calculate the average, which we refer to as reference gene-dependent QN (RQN).

*Normalization using internal control genes (NICG)[60]*

Endogenous control normalization is a technique widely used in gene expression data analysis. Its core idea is to use a set of expression-consistent genes (i.e., endogenous genes or "housekeeping genes") that are expressed at the same level in various biological conditions as reference genes to normalize the expression of other genes. The average expression of the endogenous genes in each sample is used as a normalization factor. The expression of all genes in each sample is then divided by the normalization factor for that sample, thus accounting for technical differences between samples and improving the comparability and reliability of data among various samples.

*LOG-QN*

LOG-QN will further do QN on the LOG-processed data.

*LOG-QN-Z*

LOG-QN-Z will further do QN on the LOG-processed data before doing a Z transformation.

*LOG-NPN-Z*

LOG-NPN-Z will further do NPN on the LOG-processed data before doing a Z transformation.

*LOG-NICG-Z*

LOG-NICG will further do NICG on the LOG-processed data before doing a Z transformation.

## Machine learning Models

Based on different training sets, we trained five common classifiers based on common Machine Learning algorithms: Multilayer Perceptron (MLP), Extreme Gradient Boosting (XGBoost), Logistic Regression (LR), linear Support Vector Machine (SVM), and Random Forest (RF). The five classification models presented here are all commonly used in practice, but each has different characteristics that make them suitable for comparing the interaction between dataset characteristics and models.

SVM[61] is a supervised learning algorithm that classifies data by finding the optimal hyperplane. It can be used for nonlinear problems by applying kernel tricks. SVM is particularly suitable for classification of small and medium-sized complex datasets, and handles high-dimensional data and nonlinear problems well.

LR[62] is a linear model that effectively reduces the complexity of the model and the risk of overfitting by introducing L1 regularization for feature selection. It is suitable for datasets with a large number of irrelevant features because it can help select the most useful features through sparse solution, thus improving the generalization ability of the model.

RF[63] is an integrated decision tree-based learning model that enhances the generalization ability of the model by introducing random feature selection. It is particularly effective for datasets with nonlinear, outliers and complex interactions between features.

XGBoost[64,65], which shares similarities with RF, is a high-performance model based on gradient boosting decision trees. It optimizes the regularization of the model and effectively prevents overfitting. It is ideally suited for sparse data and excels in both classification and regression problems. It performs particularly well with structured datasets.

In contrast, MLP[66,67] is a basic deep learning model containing one or more hidden layers. It is well-suited to the approximation of complex functions in pattern recognition and classification tasks, and exhibits robust learning capabilities for nonlinear relationships and highly complex patterns in data.

## Classification Performance Evaluation

Due to the multi-class and unbalanced nature of the data in this study, a combination of balanced accuracy and the Kappa statistic, in addition to F1, AUC, sensitivity, and specificity, was primarily used to evaluate classification performance based on the test set.[24,57,68-71]

The kappa statistic (Cohen's kappa) is a measure of classification accuracy that takes into account unbalanced data and chance agreement. The kappa is a statistic that compares the observed accuracy with the performance of a random classifier. It is calculated as Equation 2.

$$K=(P_o-P_e) / (1- P_e), \qquad (2)$$

where $P_o$ is the observed agreement (actual accuracy) and $P_e$ is the expected agreement under random classification. The kappa value typically ranges from -1 to 1, with 0 denoting random accuracy and 1 denoting perfect agreement.

Balanced accuracy is a metric that accounts for class imbalance and represents the average accuracy for each class. In the case of an unbalanced dataset, the overall accuracy may be high, despite the fact that the predictions for a few classes may be inaccurate. Balanced accuracy provides a fairer assessment of the model's performance across all classes. It is calculated as:

$$\text{Balanced Accuracy} = (1/n) \sum_{i=1}^{n} \left(\frac{\text{True Positives i}}{\text{Total Class i}}\right), \qquad (3)$$

where n is the number of classes.

As a typical genetic dataset, BRCA is an unbalanced dataset. Using only traditional accuracy tends to overemphasize the impact of dominant categories. The Kappa value is a measure of agreement between observed and randomized accuracy, so randomized accuracy is considered in categorical accuracy. Instead of simply calculating the total percentage of correct classifications, Balanced Accuracy is the average of the recall (or true rate) of all categories. This ensures that all categories are equally important regardless of size, thus providing a score for classifiers that performs fairly on each category. Consequently, the Kappa value is more appropriate for scenarios where random guessing performance needs to be considered, whereas Balanced Accuracy is more suitable for datasets with an imbalanced distribution, where each category must be of equal importance. The combination of balanced accuracy and Kappa value provides a more balanced and accurate assessment of model performance across all categories. In this way, any potential bias in favor of a particular category can be identified.

Based on the combination of balanced accuracy and Kappa value, we design the formula shown in Equation 4 to calculate the model evaluation value (E-value) for model selection.

$$E_{value} = -100(Kappa * Balanced\ Accuracy) * \log(\sigma_{Kappa} * \sigma_{Balanced\ Accuracy}), \qquad (4)$$

where $\sigma_{Kappa}$ and $\sigma_{Balanced\ Accuracy}$ are the variance of the corresponding Kappa and Balanced Accuracy obtained from multiple repetitions of the experiment, respectively, which can measure the robustness of the model. A large E-value corresponds to a better model performance.

# Results

We repeated the processing flow in ten times to obtain average performance metrics (**Figure 1**, **Figure 2** and **Supplementary Tables 1-2**). Regardless of the perspective, the model classification performance obtained in Model-S mode is generally better than that obtained in Model-A mode, which stems from some technical methodological, data characterization, and application differences between the datasets obtained by the two platforms. RNA-seq provides more comprehensive and precise transcriptome information.

Examining the performance metrics table corresponding to Model-A or Model-S in **Supplementary Table 1 and Table 2** separately, we find that the classification performance metrics show different trends with the changes of DEG or NDEG genes, regardless of whether we observe the performance of different classifiers under the same normalization method or the performance of different normalization methods under the same classifier. This suggests that gene selection, normalization methods and supervised machine learning classifiers need to be analyzed together.

## Results on the Raw Data

First, we consider the case where no gene selection strategy is employed. All expression values corresponding to the 15,672 genes shared by the two data platforms are directly used for the analysis to observe the performance of the five classifiers in the raw data or the data processed by different normalization methods. The performance results (**Figure 3**) show that the five different classification models present completely different patterns of change on different data, and even the classifiers do not work at all in some cases. For example, in Model-A, MLP and LR have almost no effect on raw-data, and the corresponding kappa value is close to 0.

Although MLP, SVM (Model-S) or XGB (Model-A) perform better than the others in general, and especially the models sometimes show some classification improvement on data processed by the NPN, QN, and NST normalization methods, from the point of view of practical application, both the kappa and the balanced accuracy are not satisfactory. Among them, the E-value is calculated according to Equation 4 to evaluate the model performance. XGB performs the best with a BA of 0.496 and a kappa of 0.372 when classifying is done directly on the raw data under the Model-S approach. The best performance is the combination of QN and SVM with a BA of 0.644 and a kappa of 0.460 when classified after the normalization process. The best performance of RF on the raw data was achieved with a BA of 0.389 and a Kappa of 0.352 in the Model-A approach. After normalization and then classification, the best performance was achieved with the combination of NST and XGB with a BA of 0.571 and a Kappa of 0.560.

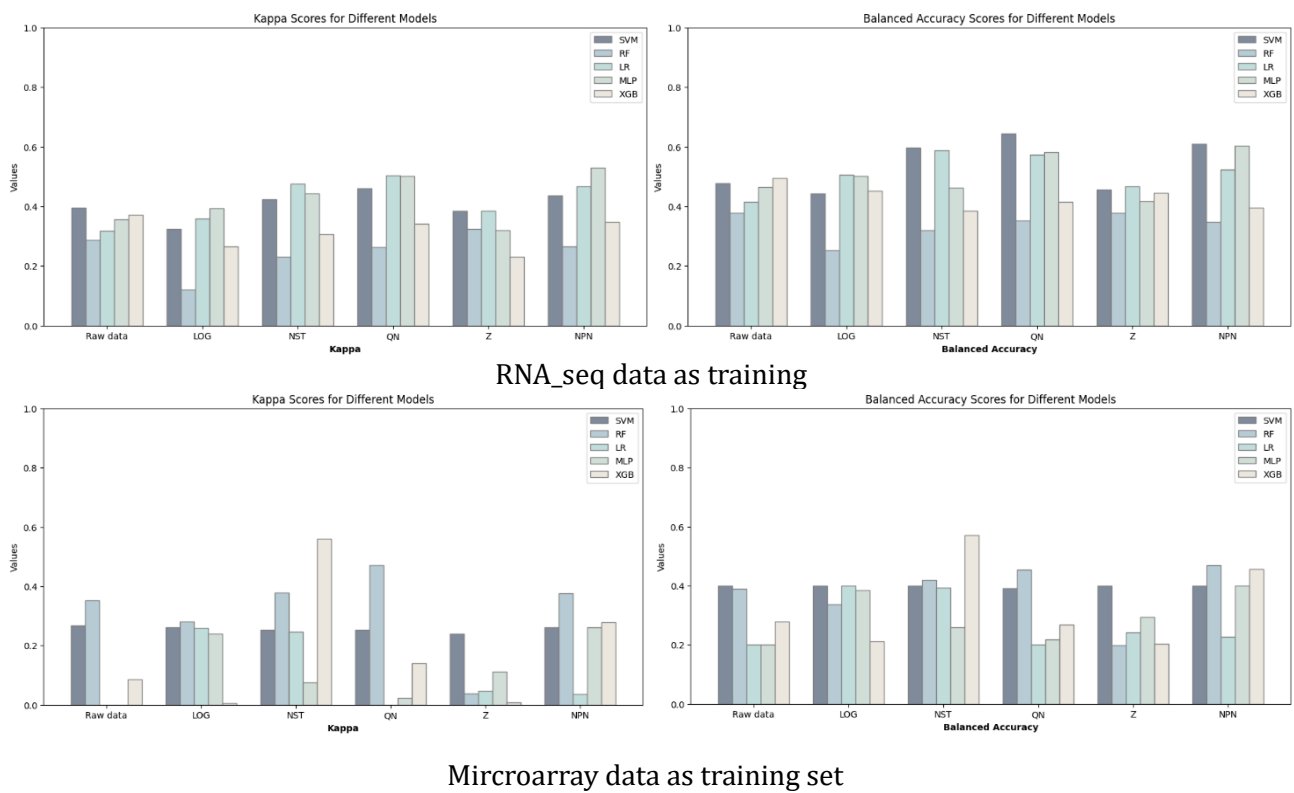

RNA_seq data as training

Mircroarray data as training set

Figure 3 The classification performance results obtained on original data

## Results on Data Selected by DEG Genes

Next, we used the gene selection strategy described above to select the expression data corresponding to the DEG genomes with P-values below a certain threshold, and analyzed the data after normalization with LOG, NPN, QN, Z and NST, respectively. The DEG gene selection threshold was varied gradually from 0.001 to 0.1, and the performance of the five classifiers on the data selected for these thresholds was observed. The obtained model evaluation metrics are shown in the **Supplementary Table1-2**. The model classification results obtained at different thresholds were compared, where the optimal performance is shown in **Figure 4**.

First of all, with the data in the **Supplementary Table 1-2**, we can observe that the classification performance of the five classifiers does not show a monotonous upward or downward trend with increasing DEG genome thresholds for data processed by any of the normalization methods. In the vast majority of cases, the classification results are not satisfactory. Compared to the case where normalization and classification are done directly on the raw data, most of the cases do not

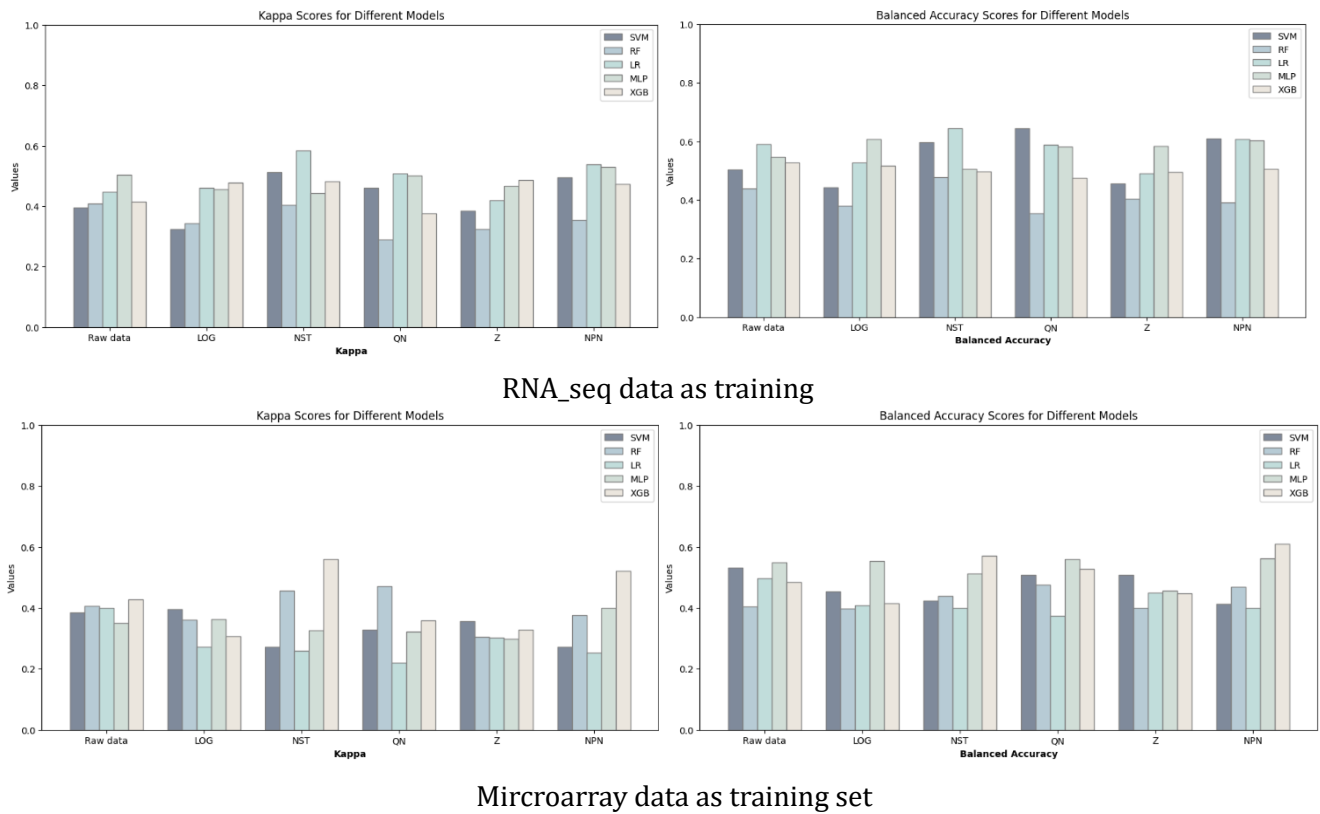

RNA_seq data as training

Mircroarray data as training set

Figure 4 The best classification performance results obtained on data selected by DEG genes. Multilayer Perceptron (MLP), Extreme Gradient Boosting (XGBoost), Logistic Regression (LR), linear Support Vector Machine (SVM), and Random Forest (RF)

show any improvement, and in some cases, the classifier does not work at all (kappa values are 0 or even negative).

Second, we also found that even when randomly dividing the training data and the test set according to the proportion of each category in the original dataset, the imbalance of the samples can lead to very different results in the repeated experiments. For example, when using MLP as a classifier with the DEG genome selection threshold set to 0.03 and using Z as the normalization method, the highest accuracy is 0.70415 and the lowest accuracy is 0.4623 in ten repeated experiments, which indicates that repartitioning leads to changes in the data, resulting in large fluctuation in the models' performance, which implies the model is less robust. This is exactly the reason why we designed the E-value that combines the mean and the variance of the Kappa and Balance Accuracy obtained from several repetitive experiments when selecting the model based on the evaluation metrics.

For the datasets selected from the DEG genome with different thresholds, the results of the statistical analysis of the performance of the different classifiers on each dataset with different normalization treatments, including the maximum value, the average performance, the standard deviation and the coefficient of variation, are shown in **Table 1**.

Looking at the performance of each classifier alone, the SVM in the results of Model A is overall more stable, as reflected by its smaller standard deviation and coefficient of variation. The other classifiers, on the other hand, show large fluctuations with the change of gene selection thresholds, and such fluctuations are not consistent across the data processed by various normalization methods. For example, the MLP classifier fluctuates more in raw and NST processed data, while the RF and XGB models fluctuate more in QN processed data. In terms of the classification performance of each classifier, there is essentially no significant improvement in classification compared to when the gene selection strategy is not used. On the contrary, when all gene data are involved in model training, better results may be achieved due to comprehensive information. In the results for the Model-S, the performance of each classifier fluctuates dramatically with the threshold, and although the SVM and MLP are slightly better overall, there is also no significant improvement in the classification performance compared to when the gene selection strategy is not used. This suggests that considering only the normalization method and gene selection strategy used for the classification model does not play a key role in the overall performance improvement.

**Table 1.** The results of the statistical analysis of the performance of the different classifiers on each dataset processed with various DEG genomes and different normalization treatments (for Model-A)

| kappa | | | | BA (balanced accuracy) | | | |
|---|---|---|---|---|---|---|---|
| SVM | Mean | Standard Deviation | Coefficient of Variation | SVM | Mean | Standard Deviation | Coefficient of Variation |
| seq_p_raw-data | 0.257 | 0.037 | 0.143 | seq_p_raw-data | 0.402 | 0.035 | 0.088 |
| seq_p_log | 0.259 | 0.038 | 0.148 | seq_p_log | 0.402 | 0.019 | 0.048 |
| seq_p_nst | 0.249 | 0.012 | 0.046 | seq_p_nst | 0.397 | 0.011 | 0.026 |
| seq_p_qn | 0.257 | 0.028 | 0.110 | seq_p_qn | 0.411 | 0.036 | 0.087 |
| seq_p_z | 0.263 | 0.035 | 0.133 | seq_p_z | 0.407 | 0.030 | 0.073 |
| seq_p_npn | 0.249 | 0.012 | 0.047 | seq_p_npn | 0.397 | 0.008 | 0.020 |
| | | | | | | | |
| RF | | | | RF | | | |
| seq_p_raw-data | 0.149 | 0.130 | 0.874 | seq_p_raw-data | 0.283 | 0.075 | 0.263 |
| seq_p_log | 0.179 | 0.106 | 0.595 | seq_p_log | 0.289 | 0.055 | 0.189 |
| seq_p_nst | 0.258 | 0.113 | 0.437 | seq_p_nst | 0.355 | 0.061 | 0.172 |
| seq_p_qn | 0.167 | 0.151 | 0.908 | seq_p_qn | 0.297 | 0.096 | 0.322 |
| seq_p_z | 0.149 | 0.101 | 0.679 | seq_p_z | 0.278 | 0.060 | 0.218 |
| seq_p_npn | 0.181 | 0.120 | 0.665 | seq_p_npn | 0.327 | 0.082 | 0.250 |
| | | | | | | | |
| LR | | | | LR | | | |
| seq_p_raw-data | 0.113 | 0.115 | 1.018 | seq_p_raw-data | 0.288 | 0.088 | 0.305 |
| seq_p_log | 0.131 | 0.098 | 0.751 | seq_p_log | 0.304 | 0.078 | 0.255 |
| seq_p_nst | 0.130 | 0.086 | 0.665 | seq_p_nst | 0.302 | 0.068 | 0.224 |
| seq_p_qn | 0.077 | 0.085 | 1.095 | seq_p_qn | 0.263 | 0.068 | 0.261 |
| seq_p_z | 0.153 | 0.095 | 0.621 | seq_p_z | 0.322 | 0.077 | 0.238 |
| seq_p_npn | 0.152 | 0.084 | 0.556 | seq_p_npn | 0.319 | 0.067 | 0.209 |
| | | | | | | | |
| MLP | | | | MLP | | | |
| seq_p_raw-data | 0.158 | 0.116 | 0.735 | seq_p_raw-data | 0.336 | 0.110 | 0.328 |
| seq_p_log | 0.202 | 0.109 | 0.540 | seq_p_log | 0.376 | 0.107 | 0.285 |
| seq_p_nst | 0.222 | 0.083 | 0.374 | seq_p_nst | 0.382 | 0.077 | 0.201 |
| seq_p_qn | 0.167 | 0.096 | 0.577 | seq_p_qn | 0.343 | 0.091 | 0.265 |
| seq_p_z | 0.181 | 0.077 | 0.424 | seq_p_z | 0.347 | 0.065 | 0.187 |
| seq_p_npn | 0.182 | 0.126 | 0.691 | seq_p_npn | 0.351 | 0.118 | 0.336 |
| | | | | | | | |
| XGB | | | | XGB | | | |
| seq_p_raw-data | 0.217 | 0.108 | 0.497 | seq_p_raw-data | 0.370 | 0.066 | 0.178 |
| seq_p_log | 0.112 | 0.105 | 0.931 | seq_p_log | 0.272 | 0.085 | 0.314 |
| seq_p_nst | 0.228 | 0.168 | 0.734 | seq_p_nst | 0.393 | 0.109 | 0.277 |
| seq_p_qn | 0.119 | 0.100 | 0.843 | seq_p_qn | 0.297 | 0.116 | 0.392 |
| seq_p_z | 0.148 | 0.098 | 0.665 | seq_p_z | 0.321 | 0.069 | 0.215 |
| seq_p_npn | 0.296 | 0.129 | 0.436 | seq_p_npn | 0.455 | 0.095 | 0.208 |

# Results on Data Selected by NDEG and DEG

Subsequently, we used a gene selection strategy to select NDEG genomes with P values above a certain threshold. Four reference gene-based normalization methods, including LOG-NPN-Z, LOG-QN, LOG-QN-Z, and LOG-NICG-Z, were used to process the corresponding gene expression data jointly selected from the NDEG genome and the DEG genome, including the training data and the test set, and then used the five classification models mentioned above to perform classification training and testing. The obtained model classification performance results are shown in the **Supplementary Tables 3-6**.

Compared with the classification results on data selected using NDEG and DEG genes, we noted the following findings (**Figure 5**).

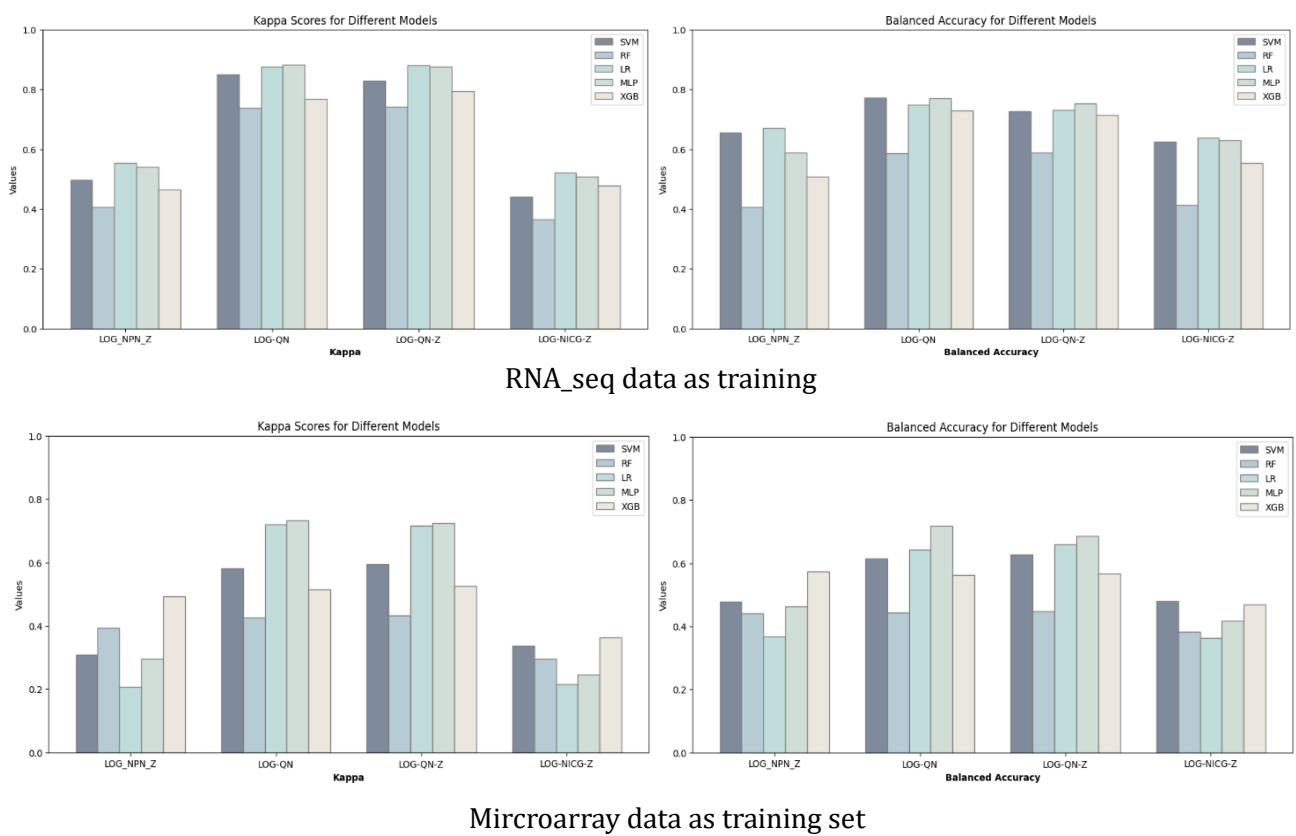

RNA_seq data as training

Mircroarray data as training set

Figure 5  The best classification performance results obtained on data selected by DEG and NDEG genes

First, for Model-S, using the data under the action of LOG-QN or LOG-QN-Z, MLP, LR and SVM classifiers can significantly improve the classification performance within the range of NDEG and DEG genome thresholds we set. Among them, MLP has a Kappa average of over 0.83 and an accuracy mean of 0.700. The classification performance of the XGB, though fluctuating greatly, is occasionally better. Further observation of the model's performance on the LOG-QN and LOG-QN-Z processed data also reveals that the surfaces corresponding to each of the key metrics in the categorization performance fluctuate considerably as the NDEG genes or DEG gene thresholds are altered, with peaks occurring at very different locations. **Table 2** shows the performance matrices obtained for Model-S by the MLP classifier in after LOG-QN normalization and for the SVM classifier after LOG-QN-Z normalization, respectively. It is important to note that the NDEG genes or DEG gene thresholds change steps are not consistent here, which are labeled in green in Table 2. For example, the MLP classifier reaches a maximum classification Balanced Accuracy of 0.771 at a NDEG gene threshold of 0.98 and a DEG gene threshold of 0.07, a maximum classification Kappa of 0.883 at a NDEG gene threshold of 0.90 and a DEG gene threshold of 0.003. The SVM classifier reaches a maximum classification Balanced Accuracy of 0.773 at a NDEG gene threshold of 0.90 and a DEG gene threshold of 0.005, a maximum classification Kappa of 0.829 at a NDEG gene threshold of 0.98 and a DEG gene threshold of 0.008. This indicates that it is more reasonable to determine the optimal model based on the model performance matrices obtained from the NDEG and DEG gene threshold changes. The optimal values in each matrix showed that the performance of both Model-S and Model-A was significantly improved **(Figure 5)**.

Second, for Model-A, the results are basically similar. MLP, LR and SVM classifiers perform better on data processed with LOG-QN or LOG-QN-Z, but the RF performance is poorer, even worse than the case without the NDEG group. The overall effect of MLP is relatively better and more stable, with the highest kappa value of 0.734 and the highest Balanced Accuracy of 0.718, and the fluctuation of the classification effect with the change of NDEG gene thresholds and the change of DEG gene thresholds is not large (SD less than 0.04). The effect of LR is more stable, but the optimal performance is not as prominent as that of MLP. The fluctuation of SVM is relatively large, and the SD value seems to be greater than 0.6. For the data under the action of LOG-NPN-Z and LOG-NICG-Z, the overall effect is unsatisfactory, in which XGB outperforms the others.

Third, regarding the normalization method, for the data under the effect of LOG-NPN-Z and LOG-NICG-Z, the no better results could be achieved regardless of the classification model used. In contrast, the performance of data under the effect of LOG-QN and LOG-QN-Zhas obvious advantages.

**Table2.** Some classification performance results on data Selected by NDEG and DEG genes (for Model-S)

| Balanced Accuracy | | | | | | | | | | | | | | | | |
|---|---|---|---|---|---|---|---|---|---|---|---|---|---|---|---|---|
| *P* values | | | | | | | | | | | | | | | | |
| MLP | | | | | | | | | | | | | | | | |
| LOG-QN | 0.001 | 0.002 | 0.003 | 0.004 | 0.005 | 0.006 | 0.007 | 0.008 | 0.009 | 0.01 | 0.02 | 0.03 | 0.05 | 0.07 | 0.1 | 1 |
| 0.98 | 0.691 | 0.698 | 0.681 | 0.705 | 0.685 | 0.702 | 0.694 | 0.687 | 0.700 | 0.692 | 0.679 | 0.685 | 0.681 | 0.771 | 0.680 | 0.702 |
| 0.95 | 0.691 | 0.721 | 0.706 | 0.694 | 0.748 | 0.695 | 0.727 | 0.671 | 0.696 | 0.687 | 0.693 | 0.694 | 0.688 | 0.689 | 0.707 | 0.705 |
| 0.92 | 0.694 | 0.675 | 0.692 | 0.686 | 0.690 | 0.702 | 0.730 | 0.704 | 0.688 | 0.684 | 0.712 | 0.664 | 0.678 | 0.711 | 0.684 | 0.706 |

| | 0.90 | 0.691 | 0.712 | 0.739 | 0.696 | 0.697 | 0.703 | 0.707 | 0.673 | 0.704 | 0.681 | 0.688 | 0.721 | 0.695 | 0.699 | 0.659 | 0.715 |
| --- | --- | --- | --- | --- | --- | --- | --- | --- | --- | --- | --- | --- | --- | --- | --- | --- | --- |
| | 0.85 | 0.690 | 0.699 | 0.691 | 0.699 | 0.676 | 0.689 | 0.699 | 0.666 | 0.715 | 0.703 | 0.674 | 0.694 | 0.682 | 0.691 | 0.694 | 0.681 |
| SVM | | | | | | | | | | | | | | | | | | |
| LOG-QN-Z | | 0.001 | 0.002 | 0.003 | 0.004 | 0.005 | 0.006 | 0.007 | 0.008 | 0.009 | 0.01 | 0.02 | 0.03 | 0.05 | 0.07 | 0.1 | 1 |
| | 0.98 | 0.738 | 0.670 | 0.668 | 0.700 | 0.735 | 0.689 | 0.702 | 0.676 | 0.667 | 0.660 | 0.604 | 0.671 | 0.656 | 0.692 | 0.648 | 0.695 |
| | 0.95 | 0.667 | 0.660 | 0.699 | 0.693 | 0.718 | 0.675 | 0.710 | 0.680 | 0.667 | 0.653 | 0.657 | 0.675 | 0.681 | 0.685 | 0.686 | 0.709 |
| | 0.92 | 0.602 | 0.634 | 0.659 | 0.713 | 0.670 | 0.693 | 0.695 | 0.682 | 0.680 | 0.674 | 0.674 | 0.597 | 0.616 | 0.657 | 0.663 | 0.681 |
| | 0.90 | 0.652 | 0.635 | 0.659 | 0.718 | 0.773 | 0.749 | 0.675 | 0.672 | 0.687 | 0.601 | 0.681 | 0.617 | 0.560 | 0.597 | 0.629 | 0.689 |
| | 0.85 | 0.587 | 0.584 | 0.618 | 0.630 | 0.645 | 0.676 | 0.434 | 0.504 | 0.612 | 0.467 | 0.431 | 0.430 | 0.547 | 0.544 | 0.548 | 0.553 |
| Kappa | | | | | | | | | | | | | | | | | | |
| MLP | | | | | | | | | | | | | | | | | | |
| LOG-QN | | 0.001 | 0.002 | 0.003 | 0.004 | 0.005 | 0.006 | 0.007 | 0.008 | 0.009 | 0.01 | 0.02 | 0.03 | 0.05 | 0.07 | 0.1 | 1 |
| | 0.98 | 0.834 | 0.827 | 0.839 | 0.852 | 0.833 | 0.845 | 0.831 | 0.809 | 0.827 | 0.845 | 0.818 | 0.825 | 0.835 | 0.850 | 0.807 | 0.838 |
| | 0.95 | 0.821 | 0.863 | 0.854 | 0.834 | 0.841 | 0.831 | 0.867 | 0.799 | 0.807 | 0.815 | 0.837 | 0.833 | 0.832 | 0.830 | 0.816 | 0.843 |
| | 0.92 | 0.827 | 0.792 | 0.837 | 0.821 | 0.837 | 0.832 | 0.872 | 0.849 | 0.821 | 0.804 | 0.857 | 0.796 | 0.808 | 0.858 | 0.823 | 0.842 |
| | 0.90 | 0.822 | 0.825 | 0.883 | 0.840 | 0.849 | 0.834 | 0.815 | 0.791 | 0.822 | 0.803 | 0.823 | 0.838 | 0.835 | 0.847 | 0.789 | 0.856 |
| | 0.85 | 0.821 | 0.844 | 0.835 | 0.852 | 0.826 | 0.830 | 0.828 | 0.786 | 0.818 | 0.825 | 0.813 | 0.815 | 0.822 | 0.851 | 0.821 | 0.822 |
| SVM | | | | | | | | | | | | | | | | | | |
| LOG-QN-Z | | 0.001 | 0.002 | 0.003 | 0.004 | 0.005 | 0.006 | 0.007 | 0.008 | 0.009 | 0.01 | 0.02 | 0.03 | 0.05 | 0.07 | 0.1 | 1 |
| | 0.98 | 0.788 | 0.808 | 0.801 | 0.783 | 0.810 | 0.811 | 0.808 | 0.829 | 0.778 | 0.793 | 0.682 | 0.787 | 0.721 | 0.818 | 0.785 | 0.797 |
| | 0.95 | 0.737 | 0.758 | 0.807 | 0.797 | 0.772 | 0.759 | 0.818 | 0.782 | 0.728 | 0.745 | 0.767 | 0.772 | 0.739 | 0.751 | 0.796 | 0.820 |
| | 0.92 | 0.697 | 0.739 | 0.744 | 0.755 | 0.751 | 0.806 | 0.782 | 0.762 | 0.701 | 0.736 | 0.716 | 0.659 | 0.732 | 0.751 | 0.770 | 0.751 |
| | 0.90 | 0.703 | 0.760 | 0.712 | 0.707 | 0.781 | 0.783 | 0.780 | 0.756 | 0.787 | 0.669 | 0.778 | 0.680 | 0.620 | 0.707 | 0.705 | 0.658 |
| | 0.85 | 0.690 | 0.703 | 0.639 | 0.706 | 0.726 | 0.675 | 0.579 | 0.586 | 0.703 | 0.541 | 0.470 | 0.519 | 0.688 | 0.649 | 0.638 | 0.547 |

The color shade indicates the ranking of the metric in the cell among all cells. The darkest red is the highest while the deepest blue shade is the lowest. MLP, Multilayer perceptron; SVM, Support vector machine.

# Discussion

## Comparison of Kappa and Balanced Accuracy

From the results of the analysis, it can be seen that in most cases, Balanced Accuracy and Kappa statistic can show similar trends, but sometimes not.

If the Kappa value is very low but the Balanced Accuracy is relatively high (e.g., in Model-A, when the NDEG threshold is 0.85 and the DEG threshold is 0.03, the normalization method is LOG-QN, and XGB is the classification model, the Kappa value obtained averages 0.354, and the balanced accuracy averages 0.531), which may indicate that although the model's performance on each category is well on average, its overall performance is not significantly improved compared to random guessing. The reason for this may be that we randomly split the training, validation, and test sets by keeping the number of samples in each of the five categories the same as the original data, which still leaves the data severely unbalanced. Once the model performs well on the main categories, which pushes up the Balanced Accuracy, the overall consistency prediction (as measured by the Kappa value) decreases due to poor performance on the categories with fewer samples. Overall, performance metrics in this case are generally not particularly impressive and the results obtained in repeated experiments vary relatively widely.

We also observed cases with high Kappa values but low balanced accuracies (e.g., in Model-S, when the NDEG threshold was 0.92, the DEG threshold was 0.03, the normalization method was LOG-QN, and the classification model was LR, the obtained Kappa values averaged 0.814 and the Balanced Accuracies averaged 0.667), which may also stem from the extreme lack of data balancing. Balanced Accuracy reflects the average of the accuracies for each category. If the model performs poorly on any of the categories, it can significantly reduce the Balanced Accuracy, which includes cases where predictions are correct on categories with small sample sizes and can be poorly predicted on major categories with large sample sizes. In this case, the overall consistency ($Po$) may still be high, and the model's overall predictions perform better compared to the random predictions, thus improving the Kappa value.

Therefore, when comparing the classification performance of different models, one cannot be limited to one performance metric. We designed an evaluation method as shown in Eq. 4. Models with high mean values and small variances of Balanced Accuracy and Kappa value obtained from multiple repetitive experiments will be regarded as well-performing models, which are more stable in their performance as well as more robust. Based on such analysis criteria, we find the optimal model. Model-S, after LOG-QN-Z processing for data with NDEG threshold of 0.85 and DEG threshold of 0.07, the MLP model outperforms the other models with balanced accuracy of 0.752 and Kappa value of 0.875. Model-A, after LOG-QN processing for data with NDEG threshold of 0.90 and DEG threshold of 0.005, the MLP model outperforms the other models with balanced accuracy of 0.707 and Kappa value of 0.734.

## Thresholds in Gene Selection Strategies

We used the F-values from the ANOVA to determine the P-values according to the F-distribution table correspondingly and used this as a threshold to achieve the selection of DEG and NDEG genes. The gene selection strategy allows for narrowing down the range of DEG genes used for classification and identifying the core NDEG genes for normalization. As the range of DEG genes is narrowed, the performance of the classification model may not improve. Even when all gene data are used for classification instead, better results can be achieved. At the same time, with the addition of NDEG genes, the classification performance improves significantly and varies with different DEG genes. This leads us to see that NDEG plays a more significant role than DEG, and that there is a large redundancy of DEG genes, but because our gene selection strategy is not good enough, we have not yet really taken out independent and representative genomes from them for classification.

Essentially, hypothesis testing is a statistical method that calculates the probability of the strength of evidence for or against the null/original hypothesis (i.e., no difference or no change) based on the sample data, which is ultimately summarized into a single value, the P value. A cut-off value (cut-off) of 0.05 and 0.95 is often chosen in various studies, which is completely arbitrary and merely an empirically generated convention. In fact, this value is not universal. For example, in disease correlation studies, perhaps a stricter cut-off value, such as 0.01, should be used. And this is exactly what our study proves. Because the thresholds of NDEG and DEG gene selection for the optimal model corresponding to Model-A are 0.90 and 0.005, respectively, and the thresholds of NDEG and DEG gene selection for the optimal model corresponding to Model-S are 0.85 and 0.07, respectively. therefore, it is necessary to find the proper thresholds on the basis of the data and the model in the course of the study.

Through the analysis, we also found that when LOG-QN or LOG-QN-Z is selected as the normalization method and MLP is selected as the classification method, the classification performance corresponding to different combinations of thresholds for NDEG and DEG shows a relatively stable effect. This suggests that under the premise of optimal selection of normalization methods and classification models, changes in the thresholds of NDEG and DEG gene selection have relatively limited effects on the final classification results. Among the three approaches of normalization method, classification model and gene selection strategy in this experiment, the normalization method and classification model currently play a more decisive role.

The amount of DEG genes selected based on P-values in our experiments is very large. First of all, the gene screening strategy in this paper only considered the variability of the features in the category and did not consider the correlation of the features, so the number of DEG genes screened was very large. In the next step, the gene screening strategy will be further improved by combining the correlation analysis. In addition, this partly stems from the high-dimensional nature of the original data itself (the number of genes is much larger than the number of samples) leading to an increased probability of false positives in the statistical test, and also stems from the possible technical variation, noise, or Batch Effect in the data, which can affect the results of the statistical test. More importantly, it reflects the skewed distribution characteristics of the data and also reaffirms that the distribution of gene expression data is usually not normally distributed. Therefore, when using traditional methods such as t-test or ANOVA, the assumption that the data conforms to a normal distribution may not be valid, leading to erroneous results. Therefore, in the future, multiple comparison corrections such as Bonferroni correction and Benjamini-Hochberg method will first be considered to control the overall false positive rate. Next, the analysis will be considered using Nonparametric statistics-based methods, and the analysis will be placed after the normalization process.

## Impact of Normalization on Models

When selecting and designing models, the potential impact of data preprocessing steps on the performance of the final model needs to be considered. Appropriate data preprocessing can improve model performance. For the BRCA of RSEM counts used in this study, we also found significant differences even when using the same classification model for data processed by different normalization methods (**Supplementary Tables**).

Comprehensively comparing the classification performance of different classification models on data processed by various normalization methods with and without NDEG, we find that the LOG-transform method and Z perform relatively poorly and QN and NPN perform relatively well when the NDEG gene is not employed, which is basically consistent with the reference 4. After adopting NDEG gene, MLP, LR and SVM achieve better performance on data processed by LOG-QN and LOG-QN-Z methods. The effect of the NDEG and DEG genes is not as prominent as that of the normalization methods. In addition, we observe that the extra added Z-transform helps in performance improvement. However, the data processed by LOG-NPN-Z and LOG-NICG-Z do not perform well in classification performance after using NDEG gene.

First, we believe that the increased span and inconsistency of cross-platform gene expression data, as well as the presence of a large amount of noise and low or extreme expression values, directly contribute to the poor performance of LOG and Z. NICG relies heavily on the stability of the selected internal control genes, which may affect the normalization results if the internal control genes are not properly selected. NST does this by converting the rankings of the data to their corresponding values in a standard normal distribution. While NPN does not assume the exact shape of the data distribution, it uses the inverse cumulative distribution function (CDF) of the target distribution to transform the quantiles values to the values of the target distribution. QN is a non-parametric normalization method that aims to make the shape of the expression distribution consistent across all samples without requiring the data to conform to a particular distribution QN also handles the sparsity of high-dimensional data more effectively than other methods. In addition, QN is better able to maintain the relative relationships between features, which is critical for certain machine learning models such as support vector machines and neural networks. In summary, the performance of the individual normalization methods is consistent with our understanding of the characteristics that gene expression data have [72]: gene expression data from microarray technology or RNA-seq sequencing usually share some common statistical properties. They often do not exactly conform to a normal distribution, but are characterized by the following non-normal distributions: high skewness, peaks in the distribution, nulls, and low expression values.

Consistent with prior reports,[14,36,73-77] our gene selection strategy is also based on ANOVA. Traditional ANOVA assumes that each group of data comes from a normal distribution. This is because the residuals (the difference between observations and group means) are assumed to be normally distributed when calculating the F-test statistic. Nonetheless, in practice, ANOVA remains robust to slight deviations from normality. When the sample size is large, the Central Limit Theorem (CLT) suggests that the distribution of the sample means tends to be normal even if the raw data does not exactly conform the normal distribution. Therefore, it is usually defaulted that ANOVA requires less normality. However, for small sample sizes, the normality assumption becomes very important. We believe this is an important reason why the gene selection strategy in this study did not play a significant role. In the future, we plan to do further research using the Non-parametric (or distribution-free) inferential statistical method to obtain more accurate core genomic information.

In summary, although all models may benefit from data normalization, different normalization treatments have different impacts on the models. Therefore, when designing and applying these models, the appropriate normalization processing method should be selected according to the specific data and the characteristics of the model to ensure that the model can perform optimally under all conditions.

## Impact Analysis of Classification Models

Different machine learning models can perform very differently on the same dataset, mainly due to the fact that each model has different learning mechanisms and approaches to processing data features. In our experimental results, we found a situation where the LR and MLP models performed the best, while the SVM performance fluctuated and the XGBoost and RF performed poorly, which we attribute to the specific characteristics of the dataset and the mechanisms by which each of these models interacts with these characteristics.

Possible factors for this phenomenon include:

1. the dimensionality and sparsity of the data

A dataset may contain many irrelevant or redundant features. LR, which implements feature selection through L1 regularization, tends to perform well on datasets with high dimensionality and low correlation between features[78]. If the dataset contains a lot of irrelevant features or noise, LR can effectively identify and compress these unimportant features to improve the model performance. MLP, on the other hand, is a powerful nonlinear model capable of capturing complex data patterns and relationships through multiple hidden layers[67]. If the feature relationships in the data are very complex and nonlinear, MLP is usually able to learn these complexities through its deep network structure.

2. Feature interaction and nonlinearity

XGBoost and RF typically perform well when feature relationships are relatively independent and linearly differentiable, and XGBoost in particular performs well for classification problems and structured datasets[65]. However, if the relationships between features in a dataset are extremely complex or masked by noise, these models may not be able to capture these relationships effectively. In particular, when gene expression data containing a large number of low or extreme values are processed more sparsely by normalization methods such as Log Transformation, Z, etc., these models may be even more unable to capture all the nonlinear patterns.

3. Model robustness and sensitivity to noise

While XGBoost and RF are resistant to general outliers and noise, they may be less effective in the face of extreme noise or outlier distributions, especially in cases where decision trees are prone to overfitting on outliers. In contrast, MLP may be better at resisting noise through its nonlinear and multilevel structure during training, especially when equipped with appropriate regularization techniques (e.g., Dropout).

4. Scale sensitivity of different models

Feature scale sensitivity is the degree to which a machine learning model is sensitive to changes in the range and scale of input feature values[79]. Different models have different sensitivities to feature scales. Distance-based models, such as LR and SVM, are very sensitive to feature scales, while tree-based models, such as decision trees, random forests, and gradient boosting trees are not sensitive to feature scales, so that the performance of the former improves much more after normalization. As a neural network model, on the other hand, the structure and learning algorithm of MLP enable it to adapt to different data scales, and the appropriate normalization method also helps to speed up the training and avoid some gradient problems, such as gradient vanishing or exploding, which leads to a more stable model performance.

By further analyzing the performance of the LR and MLP models on datasets with different preprocessing, we find that the Balanced Accuracy seems to be relatively stable with respect to the fluctuation of the Kappa value. From a data perspective, this suggests that preprocessing tools such as normalization, feature selection, and outlier handling change the distribution of the original data or the relationship between features to a certain extent, thus affecting the way the model learns. The change in data distribution directly affects the decision boundaries of the model, making the model's classification boundaries significantly different after different preprocessing, thus enhancing Kappa, which specifically emphasizes the consistency between actual and random classification. The relative stability of Balanced Accuracy, on the other hand, suggests that, despite the change in the classification boundaries, the model's ability to recognize the various categories on the whole consistency was maintained. From a modeling perspective, LR and MLP show better robustness when dealing with different data. Even if the preprocessing changes some features of the data, these two models are still able to recognize the categories effectively and maintain a more stable classification performance.

We also recognize that this study has its limitations. The training process was limited by the small number of available samples and did not take into account the effects caused by imbalance. During the analysis process, due to the limited computational power, we were unable to examine the variations in gene selection thresholds, normalization methods in a large and detailed way, especially as we mentioned earlier that further research on suitable gene selection methods is needed. However, we hope to use this study as an example to provide researchers with a comprehensive set of classification model construction strategies for various classification prediction studies.

## Conclusion

To improve ML performance in cross-platform testing on independent datasets, this study proposes a strategy based on novel NDEG-based data normalization. It combines gene selection scenarios, normalization methods and classification models. The BRCA data in TCGA were generated using both microarray and RNA-seq platforms for the sample set, and thus was used in this study. Stable NDEG and DEG with variability were first searched for by ANOVA and used for the screening of the corresponding datasets. Combining different normalization methods and classification models, different comprehensive analysis models were used for each dataset to derive classification performance metrics for cancer subtype classification. The main performance metrics, including the mean and variance of Balanced Accuracy and Kappa values, were used in the performance evaluation models we designed as a way to assess the performance of each integrated model.

The basis of machine learning model classification is first and foremost the data. In our cross-platform data classification study, RNA-seq provides more comprehensive and precise transcriptome information, so the overall performance of model S trained on RNA-seq data is much better than that of model A trained on Microarray data. The results show that NDEG and DEG gene selection can effectively improve the classification performance of the models. It is more reasonable to determine the optimal model based on the model performance matrices obtained from the NDEG and DEG gene threshold changes. The choice of normalization method is crucial for the final classification performance of the model, and the normalization methods based on parametric statistical analysis are inferior to those based on nonparametric statistics. At the same time, different classifiers perform differently on different data, and the normalization methods and classifiers should be considered together.

## Acknowledgments

## Funding

This work was supported by the U.S. National Science Foundation (IIS-2128307 to LZ) and the National Cancer Institute, National Institutes of Health (R37CA277812 to LZ).

## Author Contributions Statement

Study conceptualization and design, ensuring the data access, accuracy and integrity (LZ), and manuscript writing (DF and LZ). All authors, including DF, CHF, NG and LZ, contributed to the writing or revision of the review article and approved the final publication version.

## Conflicts of Interest

The authors declare no other conflict of interests.

## Data Availability Statement

The data sets used and/or analyzed of this study are available on the cBioPortal website (https://www.cbioportal.org/). The program coding is available from the corresponding authors on reasonable request.

## Compliance with ethical standards

This exempt study using publicly available de-identified data did not require an IRB review.

## References

1       Bhandari N, Walambe R, Kotecha K, and Khare SP. A comprehensive survey on computational learning methods for analysis of gene expression data. *Front Mol Biosci* **9**, 907150, doi:10.3389/fmolb.2022.907150 (2022).
2       Conesa A, Madrigal P, Tarazona S, Gomez-Cabrero D, Cervera A, McPherson A *et al.* A survey of best practices for RNA-seq data analysis. *Genome Biol* **17**, 13, doi:10.1186/s13059-016-0881-8 (2016).
3       Sharma A andRani R. A Systematic Review of Applications of Machine Learning in Cancer Prediction and Diagnosis. *Arch Comput Method E* **28**, 4875-4896, doi:10.1007/s11831-021-09556-z (2021).

4  Foltz SM, Greene CS, and Taroni JN. Cross-platform normalization enables machine learning model training on microarray and RNA-seq data simultaneously. *Commun Biol* **6**, 222, doi:10.1038/s42003-023-04588-6 (2023).

5  Ghandhi SA, Shuryak I, Ponnaiya B, Wu X, Garty G, Morton SR *et al.* Cross-platform validation of a mouse blood gene signature for quantitative reconstruction of radiation dose. *Sci Rep* **12**, 14124, doi:10.1038/s41598-022-18558-1 (2022).

6  Wang G, Kitaoka T, Crawford A, Mao Q, Hesketh A, Guppy FM *et al.* Cross-platform transcriptomic profiling of the response to recombinant human erythropoietin. *Sci Rep* **11**, 21705, doi:10.1038/s41598-021-00608-9 (2021).

7  Angel PW, Rajab N, Deng Y, Pacheco CM, Chen T, Le Cao KA *et al.* A simple, scalable approach to building a cross-platform transcriptome atlas. *PLoS Comput Biol* **16**, e1008219, doi:10.1371/journal.pcbi.1008219 (2020).

8  Franks JM, Cai G, and Whitfield ML. Feature specific quantile normalization enables cross-platform classification of molecular subtypes using gene expression data. *Bioinformatics* **34**, 1868-1874, doi:10.1093/bioinformatics/bty026 (2018).

9  Ritchie MD, Holzinger ER, Li R, Pendergrass SA, and Kim D. Methods of integrating data to uncover genotype-phenotype interactions. *Nat Rev Genet* **16**, 85-97, doi:10.1038/nrg3868 (2015).

10  Le Cao KA, Rohart F, McHugh L, Korn O, and Wells CA. YuGene: a simple approach to scale gene expression data derived from different platforms for integrated analyses. *Genomics* **103**, 239-251, doi:10.1016/j.ygeno.2014.03.001 (2014).

11  Pacini C, Dempster JM, Boyle I, Goncalves E, Najgebauer H, Karakoc E *et al.* Integrated cross-study datasets of genetic dependencies in cancer. *Nat Commun* **12**, 1661, doi:10.1038/s41467-021-21898-7 (2021).

12  Nam AS, Chaligne R, and Landau DA. Integrating genetic and non-genetic determinants of cancer evolution by single-cell multi-omics. *Nat Rev Genet* **22**, 3-18, doi:10.1038/s41576-020-0265-5 (2021).

13  Sharif MI, Li JP, Naz J, and Rashid I. A comprehensive review on multi-organs tumor detection based on machine learning. *Pattern Recognition Letters* **131**, 30-37 (2020).

14  Thalor A, Kumar Joon H, Singh G, Roy S, and Gupta D. Machine learning assisted analysis of breast cancer gene expression profiles reveals novel potential prognostic biomarkers for triple-negative breast cancer. *Comput Struct Biotechnol J* **20**, 1618-1631, doi:10.1016/j.csbj.2022.03.019 (2022).

15  Thompson JA, Tan J, and Greene CS. Cross-platform normalization of microarray and RNA-seq data for machine learning applications. *PeerJ* **4**, e1621, doi:10.7717/peerj.1621 (2016).

16  Majid A, Ali S, Iqbal M, and Kausar N. Prediction of human breast and colon cancers from imbalanced data using nearest neighbor and support vector machines. *Computer methods and programs in biomedicine* **113**, 792-808 (2014).

17  Kourou K, Exarchos TP, Exarchos KP, Karamouzis MV, and Fotiadis DI. Machine learning applications in cancer prognosis and prediction. *Computational and structural biotechnology journal* **13**, 8-17 (2015).

18  Maldonado S, Weber R, and Famili F. Feature selection for high-dimensional class-imbalanced data sets using Support Vector Machines. *Information Sciences* **286**, 228-246, doi:10.1016/j.ins.2014.07.015 (2014).

19  Abdulrauf Sharifai G andZainol Z. Feature Selection for High-Dimensional and Imbalanced Biomedical Data Based on Robust Correlation Based Redundancy and Binary Grasshopper Optimization Algorithm. *Genes (Basel)* **11**, doi:10.3390/genes11070717 (2020).

20  Yijing L, Haixiang G, Xiao L, Yanan L, and Jinling L. Adapted ensemble classification algorithm based on multiple classifier system and feature selection for classifying multi-class imbalanced data. *Knowledge-Based Systems* **94**, 88-104, doi:10.1016/j.knosys.2015.11.013 (2016).

21   Feng CH, Disis ML, Cheng C, and Zhang L. Multimetric feature selection for analyzing multicategory outcomes of colorectal cancer: random forest and multinomial logistic regression models. *Lab Invest* **102**, 236-244, doi:10.1038/s41374-021-00662-x (2022).
22   Hambali MA, Oladele TO, and Adewole KS. Microarray cancer feature selection: Review, challenges and research directions. *International Journal of Cognitive Computing in Engineering* **1**, 78-97, doi:10.1016/j.ijcce.2020.11.001 (2020).
23   Zheng Y, Li Y, Wang G, Chen Y, Xu Q, Fan J, and Cui X. A hybrid feature selection algorithm for microarray data. *The Journal of Supercomputing* **76**, 3494-3526, doi:10.1007/s11227-018-2640-y (2018).
24   Bajer D, Zorić B, Dudjak M, and Martinović G. in 2019 IEEE 15th International Scientific Conference on Informatics.  000285-000292 (IEEE).
25   Deng F, Zhao L, Yu N, Lin Y, and Zhang L. Union With Recursive Feature Elimination: A Feature Selection Framework to Improve the Classification Performance of Multicategory Causes of Death in Colorectal Cancer. *Lab Invest* **104**, 100320, doi:10.1016/j.labinv.2023.100320 (2024).
26   Haixiang G, Yijing L, Shang J, Mingyun G, Yuanyue H, and Bing G. Learning from class-imbalanced data: Review of methods and applications. *Expert Systems with Applications* **73**, 220-239, doi:10.1016/j.eswa.2016.12.035 (2017).
27   Bolón-Canedo V, Sánchez-Maroño N, and Alonso-Betanzos A. Feature selection for high-dimensional data. *Progress in Artificial Intelligence* **5**, 65-75, doi:10.1007/s13748-015-0080-y (2016).
28   Hira ZM andGillies DF. A Review of Feature Selection and Feature Extraction Methods Applied on Microarray Data. *Adv Bioinformatics* **2015**, 198363, doi:10.1155/2015/198363 (2015).
29   Ai C. A Method for Cancer Genomics Feature Selection Based on LASSO-RFE. *Iranian Journal of Science and Technology, Transactions A: Science* **46**, 731-738, doi:10.1007/s40995-022-01292-8 (2022).
30   Song Y, Wang Y, Geng X, Wang X, He H, Qian Y *et al.* Novel biomarker genes for the prediction of post-hepatectomy survival of patients with NAFLD-related hepatocellular carcinoma. *Cancer Cell Int* **23**, 269, doi:10.1186/s12935-023-03106-2 (2023).
31   Song R, He S, Wu Y, Chen W, Song J, Zhu Y *et al.* Validation of reference genes for the normalization of the RT-qPCR in peripheral blood mononuclear cells of septic patients. *Heliyon* **9**, e15269, doi:10.1016/j.heliyon.2023.e15269 (2023).
32   Bairakdar MD, Tewari A, and Truttmann MC. A meta-analysis of RNA-Seq studies to identify novel genes that regulate aging. *Exp Gerontol* **173**, 112107, doi:10.1016/j.exger.2023.112107 (2023).
33   Veryaskina YA, Titov SE, Ivanov MK, Ruzankin PS, Tarasenko AS, Shevchenko SP *et al.* Selection of reference genes for quantitative analysis of microRNA expression in three different types of cancer. *PLoS One* **17**, e0254304, doi:10.1371/journal.pone.0254304 (2022).
34   da Conceicao Braga L, Goncalves BOP, Coelho PL, da Silva Filho AL, and Silva LM. Identification of best housekeeping genes for the normalization of RT-qPCR in human cell lines. *Acta Histochem* **124**, 151821, doi:10.1016/j.acthis.2021.151821 (2022).
35   Echle A, Rindtorff NT, Brinker TJ, Luedde T, Pearson AT, and Kather JN. Deep learning in cancer pathology: a new generation of clinical biomarkers. *Br J Cancer* **124**, 686-696, doi:10.1038/s41416-020-01122-x (2021).
36   Bhuva DD, Cursons J, and Davis MJ. Stable gene expression for normalisation and single-sample scoring. *Nucleic Acids Res* **48**, e113, doi:10.1093/nar/gkaa802 (2020).
37   Xu L, Luo H, Wang R, Wu WW, Phue JN, Shen RF *et al.* Novel reference genes in colorectal cancer identify a distinct subset of high stage tumors and their associated histologically normal colonic tissues. *BMC Med Genet* **20**, 138, doi:10.1186/s12881-019-0867-y (2019).

38   Wang Z, Lyu Z, Pan L, Zeng G, and Randhawa P. Defining housekeeping genes suitable for RNA-seq analysis of the human allograft kidney biopsy tissue. *BMC Med Genomics* **12**, 86, doi:10.1186/s12920-019-0538-z (2019).
39   Tong L, Wu PY, Phan JH, Hassazadeh HR, Consortium S, Tong W, and Wang MD. Impact of RNA-seq data analysis algorithms on gene expression estimation and downstream prediction. *Sci Rep* **10**, 17925, doi:10.1038/s41598-020-74567-y (2020).
40   Singh D andSingh B. Investigating the impact of data normalization on classification performance. *Applied Soft Computing* **97**, doi:10.1016/j.asoc.2019.105524 (2020).
41   Jo J, Choi S, Oh J, Lee SG, Choi SY, Kim KK, and Park C. Conventionally used reference genes are not outstanding for normalization of gene expression in human cancer research. *BMC Bioinformatics* **20**, 245, doi:10.1186/s12859-019-2809-2 (2019).
42   Faraldi M, Gomarasca M, Sansoni V, Perego S, Banfi G, and Lombardi G. Normalization strategies differently affect circulating miRNA profile associated with the training status. *Sci Rep* **9**, 1584, doi:10.1038/s41598-019-38505-x (2019).
43   Evans C, Hardin J, and Stoebel DM. Selecting between-sample RNA-Seq normalization methods from the perspective of their assumptions. *Brief Bioinform* **19**, 776-792, doi:10.1093/bib/bbx008 (2018).
44   Abbas-Aghababazadeh F, Li Q, and Fridley BL. Comparison of normalization approaches for gene expression studies completed with high-throughput sequencing. *PLoS One* **13**, e0206312, doi:10.1371/journal.pone.0206312 (2018).
45   Cheng L, Lo LY, Tang NL, Wang D, and Leung KS. CrossNorm: a novel normalization strategy for microarray data in cancers. *Sci Rep* **6**, 18898, doi:10.1038/srep18898 (2016).
46   Zyprych-Walczak J, Szabelska A, Handschuh L, Gorczak K, Klamecka K, Figlerowicz M, and Siatkowski I. The Impact of Normalization Methods on RNA-Seq Data Analysis. *Biomed Res Int* **2015**, 621690, doi:10.1155/2015/621690 (2015).
47   Schwarzenbach H, da Silva AM, Calin G, and Pantel K. Data Normalization Strategies for MicroRNA Quantification. *Clin Chem* **61**, 1333-1342, doi:10.1373/clinchem.2015.239459 (2015).
48   Li P, Piao Y, Shon HS, and Ryu KH. Comparing the normalization methods for the differential analysis of Illumina high-throughput RNA-Seq data. *BMC Bioinformatics* **16**, 347, doi:10.1186/s12859-015-0778-7 (2015).
49   Risso D, Ngai J, Speed TP, and Dudoit S. Normalization of RNA-seq data using factor analysis of control genes or samples. *Nat Biotechnol* **32**, 896-902, doi:10.1038/nbt.2931 (2014).
50   Maza E, Frasse P, Senin P, Bouzayen M, and Zouine M. Comparison of normalization methods for differential gene expression analysis in RNA-Seq experiments: A matter of relative size of studied transcriptomes. *Commun Integr Biol* **6**, e25849, doi:10.4161/cib.25849 (2013).
51   Dillies MA, Rau A, Aubert J, Hennequet-Antier C, Jeanmougin M, Servant N *et al.* A comprehensive evaluation of normalization methods for Illumina high-throughput RNA sequencing data analysis. *Brief Bioinform* **14**, 671-683, doi:10.1093/bib/bbs046 (2013).
52   Li J, Witten DM, Johnstone IM, and Tibshirani R. Normalization, testing, and false discovery rate estimation for RNA-sequencing data. *Biostatistics* **13**, 523-538, doi:10.1093/biostatistics/kxr031 (2012).
53   Quackenbush J. Microarray data normalization and transformation. *Nat Genet* **32 Suppl**, 496-501, doi:10.1038/ng1032 (2002).
54   Khan Y, Hammarström D, Ellefsen S, and Ahmad R. Normalization of gene expression data revisited: the three viewpoints of the transcriptome in human skeletal muscle undergoing load-induced hypertrophy and why they matter. *BMC bioinformatics* **23**, 241 (2022).
55   Hansen KD, Irizarry RA, and Wu Z. Removing technical variability in RNA-seq data using conditional quantile normalization. *Biostatistics* **13**, 204-216, doi:10.1093/biostatistics/kxr054 (2012).

56 Kim J-H. Estimating classification error rate: Repeated cross-validation, repeated hold-out and bootstrap. *Computational Statistics & Data Analysis* **53**, 3735-3745, doi:10.1016/j.csda.2009.04.009 (2009).

57 Raschka S. Model evaluation, model selection, and algorithm selection in machine learning. arXiv 2018. *arXiv preprint arXiv:1811.12808* (2021).

58 Conover WJ, Tercero-Gómez VG, and Cordero-Franco AE. The sequential normal scores transformation. *Sequential Analysis* **36**, 397-414 (2017).

59 Brodsky E andDarkhovsky BS. Non-Parametric Statistical Diagnosis: Problems and Methods. Springer Netherlands, 2013.

60 Vandesompele J, De Preter K, Pattyn F, Poppe B, Van Roy N, De Paepe A, and Speleman F. Accurate normalization of real-time quantitative RT-PCR data by geometric averaging of multiple internal control genes. *Genome biology* **3**, 1-12 (2002).

61 Steinwart I andChristmann A. Support Vector Machines. Springer New York, 2008.

62 Hosmer DW, Lemeshow S, and Sturdivant RX. Applied Logistic Regression. Wiley, 2013.

63 Kulkarni VY andSinha PK. Random forest classifiers: a survey and future research directions. *Int. J. Adv. Comput* **36**, 1144-1153 (2013).

64 Ma B, Meng F, Yan G, Yan H, Chai B, and Song F. Diagnostic classification of cancers using extreme gradient boosting algorithm and multi-omics data. *Comput Biol Med* **121**, 103761, doi:10.1016/j.compbiomed.2020.103761 (2020).

65 Sheridan RP, Wang WM, Liaw A, Ma J, and Gifford EM. Extreme gradient boosting as a method for quantitative structure–activity relationships. *Journal of chemical information and modeling* **56**, 2353-2360 (2016).

66 Karthik S andSudha M. A survey on machine learning approaches in gene expression classification in modelling computational diagnostic system for complex diseases. *International Journal of Engineering and Advanced Technology* **8**, 182-191 (2018).

67 Dunne RA. A statistical approach to neural networks for pattern recognition. John Wiley & Sons, 2007.

68 Zhou J, Gandomi AH, Chen F, and Holzinger A. Evaluating the quality of machine learning explanations: A survey on methods and metrics. *Electronics* **10**, 593 (2021).

69 Handelman GS, Kok HK, Chandra RV, Razavi AH, Huang S, Brooks M *et al.* Peering into the black box of artificial intelligence: evaluation metrics of machine learning methods. *American Journal of Roentgenology* **212**, 38-43 (2019).

70 Carvalho DV, Pereira EM, and Cardoso JS. Machine learning interpretability: A survey on methods and metrics. *Electronics* **8**, 832 (2019).

71 Vujović Ž. Classification model evaluation metrics. *International Journal of Advanced Computer Science and Applications* **12**, 599-606 (2021).

72 Gentleman R, Carey V, Huber W, Irizarry R, and Dudoit S. Bioinformatics and computational biology solutions using R and Bioconductor. Springer Science & Business Media, 2005.

73 Molania R, Foroutan M, Gagnon-Bartsch JA, Gandolfo LC, Jain A, Sinha A *et al.* Removing unwanted variation from large-scale RNA sequencing data with PRPS. *Nat Biotechnol* **41**, 82-95, doi:10.1038/s41587-022-01440-w (2023).

74 Cui X andChurchill GA. Statistical tests for differential expression in cDNA microarray experiments. *Genome biology* **4**, 1-10 (2003).

75 Jiang K, Koob J, Chen XD, Krajeski RN, Zhang Y, Volf V *et al.* Programmable eukaryotic protein synthesis with RNA sensors by harnessing ADAR. *Nat Biotechnol* **41**, 698-707, doi:10.1038/s41587-022-01534-5 (2023).

76 Graf J, Cho S, McDonough E, Corwin A, Sood A, Lindner A *et al.* FLINO: a new method for immunofluorescence bioimage normalization. *Bioinformatics* **38**, 520-526, doi:10.1093/bioinformatics/btab686 (2022).

77 Lin Y, Golovnina K, Chen ZX, Lee HN, Negron YL, Sultana H *et al.* Comparison of normalization and differential expression analyses using RNA-Seq data from 726 individual Drosophila melanogaster. *BMC Genomics* **17**, 28, doi:10.1186/s12864-015-2353-z (2016).

78     Venkatesh B andAnuradha J. A review of feature selection and its methods. *Cybernetics and information technologies* **19**, 3-26 (2019).

79     Wu J, Kong L, Yi M, Chen Q, Cheng Z, Zuo H, and Yang Y. Prediction and screening model for products based on fusion regression and xgboost classification. *Computational Intelligence and Neuroscience* **2022**, 4987639 (2022).